\documentclass[useAMS,usenatbib]{mn2e}
\usepackage{amsmath, amssymb}
\usepackage{natbib}
\usepackage{amssymb}
\usepackage{graphicx}
\usepackage{multirow}
\usepackage{times}
\usepackage[normalem]{ulem}
\usepackage{color}
\newcommand{\Gb}{\Gamma_b}
\newcommand{\Gj}{\gamma_j'}

\newcommand{\tdtpar}{\left(\frac{T}{\delta t}\right)}
\newcommand{\lp}{l'}
\newcommand{\tgg}{\tau_{\gamma\gamma}}
\newcommand{\tobs}{{T_{\rm obs}}}
\newcommand{\tmax}{{T_{\rm max}}}
\newcommand{\ton}{T}

\def\gsim{ \lower .75ex \hbox{$\sim$} \llap{\raise .27ex \hbox{$>$}} }
\def\lsim{ \lower .75ex\hbox{$\sim$} \llap{\raise .27ex \hbox{$<$}} }

\title{Variability in Blazars:  Clues from PKS 2155-304}
\author[R. Narayan  and T. Piran]
{Ramesh Narayan$^1$\thanks{E-mail: rnarayan@cfa.harvard.edu (RN);
tsvi@phys.huji.ac.il (TP)}; 
Tsvi Piran$^2$\footnotemark[1]\\
$^1${Harvard-Smithsonian Center for Astrophysics, 60 Garden Street, Cambridge, MA 02138, USA}\\
$^2${Racah Institute of Physics, The Hebrew University, Jerusalem 91904, Israel}\\
}

\begin{document}
\date{\today}
\maketitle
\label{firstpage}

\begin{abstract}

Rapid variability on a time scale much faster than the light-crossing
time of the central supermassive black hole has been seen in TeV
emission from the blazar PKS 2155-304. The most plausible explanation
of this puzzling observation is that the radiating fluid in the
relativistic jet is divided into a large number of sub-regions which
move in random directions with relative Lorentz factors $\approx
\Gj$. The random motions introduce new relativistic effects, over and
above those due to the overall mean bulk Lorentz factor $\Gb$ of the
jet.  We consider two versions of this ``jets in a jet'' model.  In
the first, the ``subjets'' model, stationary regions in the mean jet
frame emit relativistic subjets that produce the observed
radiation. The variability time scale is determined by the size of the
sub-regions in the mean jet frame.  This model, which is motivated by
magnetic reconnection, has great difficulty explaining the
observations in PKS 2155-304.  In the alternate ``turbulence'' model,
various sub-regions move relativistically in random directions and the
variability time scale is determined by the size of these regions in
their own comoving frames.  This model fits the data much more
comfortably.  Details such as what generates the turbulent motion, how
particles are heated, and what the radiation process is, remain to be
worked out.  We consider collisions between TeV photons emitted from
different sub-regions and find that, in both the subjets and
turbulence models, the mean bulk Lorentz factor $\Gb$ of the jet needs
to be $>25$ to avoid the pair catastrophe.

\end{abstract}

\begin{keywords}
black hole physics -- magnetic reconnection -- relativistic processes
-- turbulence -- galaxies: BL Lacertae objects: individual: PKS
2155-304 -- galaxies: jets
\end{keywords}

\section{Introduction}

On July 28, 2006, the high-frequency peaked BL Lac source PKS 2155-304
had a strong flare in the TeV band, with an average flux that was more
than 10 times larger than the typical flux seen at other times
\citep{Aharonian+07}. During this flare, which lasted for about an
hour, rapid variability on a time scale $\approx 300$\,s was observed.
With a galaxy bulge luminosity $M_R = -24.4 $ \citep{Kotilainen+98},
the expected mass of the black hole (BH) in the nucleus of the galaxy
is $M_{\rm BH} \approx 1-2 \times 10^9 M_\odot$
\citep{Bettoni+03}. This corresponds to a gravitational radius $R_g
\equiv 2GM/c^2 \approx 3-6 \times 10^{14}$\,cm, and a corresponding
time scale $T_g=R_g/c \approx1-2 \times 10^4$\,s.  Amazingly, $T_g$ is
nearly two orders of magnitude longer than the variability time.

It is generally accepted that TeV emission in blazars comes from
relativistic jets that are pointed towards the observer
\citep{Hinton_Hofmann09}. Opacity arguments suggest that the bulk
Lorentz factor of the jet in PKS 2155-304 must be very large, $\Gb>50$
\citep{Begelman+08}. At first sight, it might appear that this large
Lorentz factor will also explain the rapid variability observed in the
source.  However, while relativity can certainly cause the observed
variability time to be shorter than $R/c$, where $R$ is the size of
the emitting region, there is no simple way for the variability time
to be shorter than the gravitational time scale $T_g$ of the central
engine. In PKS 2155-304, the variability is faster by a factor of
several tens.

The only way to understand the rapid variability in PKS 2155-304
is if one of the following conditions holds: (i) the entire source
(engine and emitting region) moves towards the observer with a Lorentz
factor $\approx 50$, or (ii) the emitting region alone moves rapidly
towards the observer, and the variability is caused by some local
instability in the radiating gas which is insensitive to any time
scale associated with the central BH engine, or (iii) the supermassive
BH is $\approx 50$ times less massive than we estimate.  The first
option is obviously impossible. We do not expect a $10^9 M_\odot$ BH
to move with a Lorentz factor of 50.  The third option is also
unlikely since the BH mass estimate is obtained via the well-known (BH
mass)$-$(bulge luminosity) relation \citep{Magorrian+98}, which has an
uncertainty of no more than a factor of a few\footnote{An exception
would arise, of course, if the system consists of a binary black hole and
the smaller BH produces the flare \citep{VolpeRieger}.}.  We are thus
led to the second option, but even this is highly non-trivial.  The
size of the emitting region is surely at least as large as $R_g$,
which means, given the short variability time scale, that only a small
fraction of the emitting volume can be involved in the
variability. Yet, the large amplitude of the observed fluctuations
suggests that the whole emitting region must contribute to the
flare. There is an inherent paradox in these conflicting indications.

Building on ideas developed earlier to explain variability in
gamma-ray bursts (GRBs, see 
\citealt{Lyutikov_Blandford03,Lyutikov06,Narayan_Kumar09,Lazar+09}),
Giannios et al. (2009; see also \citealt{Giannios+10,Nalewajko+11})
proposed a ``jets in a jet'' model as a possible
explanation of the TeV flare observed in PKS 2155-304.  According to
this model, a relativistic jet moves with an overall bulk Lorentz
factor $\Gb$ towards the observer, but the TeV emission is
produced in emitting regions that themseves move relativistically with
respect to the mean frame of the jet.  As shown in the context of GRBs
\citep{Narayan_Kumar09,Lazar+09}, such a model naturally produces
large amplitude variability on time scales much shorter than the
light-crossing time of the emitting volume.

In this work we explore the ``jets in a jet'' model and obtain several
constraints that must be satisfied for this model to operate.  Our
analysis follows the work of \cite{Narayan_Kumar09} and
\cite{Lazar+09} on the equivalent problem for GRBs.  In \S 2 we
summarize the relevant observations of PKS 2155-304. In \S 3 we
discuss the overall issue of time scales in relativistic jets. We then
describe the ``jets in a jet" model and its two basic variants ---
``subjets" (\S3.1), ``turbulence" (\S3.2)
--- and analyze the conditions under which each can fit the
observations. Both variants of the model involve a large number of
emitting regions moving in different directions. Photons emitted in
one region can interact with those emitted in another, and if the
optical depth is sufficiently large, colliding high energy photons
will pair produce and make the source opaque. We examine the optical
depth to this process in \S 5 and consider how the constraint $\Gb>50$
obtained by \citet{Begelman+08} is modified in the ``jets in a jet"
scenario.  In \S 6 we summarize our results and consider the
implications. Additional details are given in two Appendices.

\section{Observations}
\label{obs}

The TeV flare under consideration in PKS 2155-304 was already in
progress when observations began on July 28, 2006
\citep{Aharonian+07}. It lasted for about $T_{\rm obs}\approx
4000$\,s, after which the flux fell to a much lower value.  For the
analysis in this paper, we need to know the total duration $T$ of the
flare.  In Appendix A, we use a Bayesian analysis to estimate the
probability distribution of $T$.  We estimate the median duration of
the flare to be $T_{\rm median}\approx 2T_{\rm obs}\approx 8000$\,s,
and the mean duration to be $T_{\rm mean}=\ln(T_{\rm max}/T_{\rm
obs})\,T_{\rm obs} \approx (3-10)\,\tobs \approx 12000-40000$\,s.  For the
present analysis, we choose $T \approx 20000$\,s (corresponding to
$T_{\rm max}\approx$ a week).

The second quantity we need is the variability time scale.
\citet{Aharonian+07} found very rapid rise times
$\tau_r\sim100-200$\,s and longer decay times $\tau_d\sim200-600$\,s
(see their Table 1). They also defined a doubling time $T_2$, and
state that the fastest $T_2=224\pm60$\,s and the average
$T_2=330\pm40$\,s.  Based on this information, we choose the
characteristic variability time to be $\delta t \approx 300$\,s (as in
\citealt{Begelman+08}).
The ratio of the total duration of the flare to the variability time
is an important quantity\footnote{This quantity is sometimes called
the source variability $V$ in the GRB literature
\citep{Sari_Piran97,Narayan_Kumar09}}. For the above choices, we find
\begin{equation}
\frac{T}{\delta t} \approx 70, \qquad T \approx
20000\,{\rm s}, \qquad \delta t \approx 300\,{\rm s}.
\label{eq:V}
\end{equation}

During the $4000$\,s of observations of the flare in PKS 2155-304,
there were 5 major pulses in the TeV emission. Scaling this number to
the total duration $T$, we estimate the total number of pulses in the
flare to be $N_p \approx 25$.  From this we estimate the duty cycle
$\xi$ of the pulsed emission:
\begin{equation}
\xi \equiv \frac{N_p \delta t}{T} \approx 0.4, \qquad N_p \approx 25.
\label{eq:np}
\end{equation}
The numerical values given in equations (\ref{eq:V}) and (\ref{eq:np})
are for our fiducial estimates of $T$ and $\delta t$. As a very
extreme case, we also sometimes consider $T=\tobs \approx 4000$\,s,
$\delta t \approx 600$\,s, which give $T/\delta t\approx 7$, $\xi
\approx 0.8$.

As mentioned earlier, the mass of the BH in PKS 2155-304 is estimated
to be $M_{\rm BH} \approx 1-2 \times 10^9 M_\odot$, so its
gravitational time scale is
\begin{equation}
T_g \approx 10000 M_9\,{\rm s}, \qquad M_9 \equiv \frac{M_{\rm BH}}{10^9
M_\odot} \approx 1-2.
\end{equation}
In the following we use a canonical value of $M_9=2$, such that $T_g$
matches our estimate of the flare duration $T$.  However, there is no
direct measurement of the BH mass, so we also occasionally consider a
lower value: $M_9=0.5$.

\section {The model}
\label{model}

Naively, one might think that relativistic motion can cause the
observed variability time of a source to be arbitrarily small and so
rapid variability can always be explained.  However, the situation is
more complicated.

Consider a smooth homogeneous jet moving towards the observer with a
Lorentz factor $\Gb$. Let the jet be at a distance $R$ from the
central engine and have a radial width $\Delta R$ as measured in the
``lab'' frame, i.e., the frame of the host galaxy.  The emission is
beamed into an angle $1/\Gb$, so a distant observer receives radiation
from only a region of lateral size $\sim R /\Gb$.  The angular time
scale $t_{\rm ang}$, the time delay between the center and edge of the
visible patch, is $\sim R/2 c \Gb^2$.  Since the jet is assumed
to be smooth, any observed variability can be no faster than
this.\footnote{Throughout this paper, for simplicity, we ignore the
cosmological time dilation factor $(1+z)=1.116$ between the host
galaxy of PKS 2155-304 and the observer.  This factor will be more
important for a source at a higher redshift. In that case, time scales
in the observer frame will need to be multiplied by $(1+z)$.} 

In the comoving frame of the jet, the size of a causally connected
region is $R' \sim R/\Gb$, where here and throughout the
paper we use a prime to distinguish length and time scales in the
comoving frame from those in the lab frame.  The radial width $\Delta
R'$ of the shell of material in the jet cannot be smaller than this
size.  (If it starts out smaller, it will quickly expand to this size
as a result of internal pressure gradients.)  Thus, two photons
emitted simultaneously in the fluid frame from the front and back of
the shell will reach the observer with a time difference $\Delta R'/2
c \Gb \,\gsim\, R/2 c \Gb^2$. This radial time scale $t_{\rm rad}$ is
of the same order as, or longer than, the angular time scale $t_{\rm
ang}$.

Remarkably, $R/2 c \Gb^2$ is also the observed time difference
between two photons emitted by a single fluid element, one emitted
when the fluid is at a distance $R$ from the engine and the other
emitted at a distance $2R$. This dynamical time scale $t_{\rm dyn}$ is
relevant, for instance, if the jet slows down as a result of
interacting with a surrounding medium.  It is also relevant for a
freely streaming jet, since it is the time scale on which the
energetic particles in the jet are cooled by adiabatic expansion.

For large values of $\Gb$, all three time scales described above are
much shorter than the Newtonian time scale $R/c$ that one would
compute based on the linear size of the source. Thus relativistic
motion is indeed capable of producing very short time scales compared
to the dimensions of the emitting region.

However, there is a fourth time scale.  From a simple light-crossing
argument applied to the power source of the jet, the radial width of
the emitting shell of material cannot be smaller than the size $R_g$
of the central engine.  Thus, the shell width in the lab frame must
satisfy $\Delta R \geq R_g$, or equivalently, in the jet frame $\Delta
R' \geq \Gb R_g$. This additional constraint means that the observed
variability time scale cannot be shorter than $T_g$. However, in PKS
2155-304, we have $T_g\approx 20000$\,s, whereas the variability time
is much shorter, $\delta t \approx 300$\,s.  This is the problem we
are faced with.

In order to solve the above problem, we have to give up the assumption
of a smooth jet and must allow the radiation observed at any given
time to be emitted from a tiny region of the source with a size $\lp
\ll R'$.  However, $R'$ is the size of a causally connected
region. How can a significant amount of energy, as needed to produce
the observed large amplitude pulses of TeV emission, be squeezed into
a region much smaller than $R'$?  The answer is (Narayan \& Kumar
2009; Lazar et al. 2009; Giannios et al. 2009; see Appendix B for a
Table comparing the notations used in these papers with that in the
present work) to make the additional assumption that the radiating
region under consideration moves relativistically with respect to the
comoving frame of the jet.  This causes the radiation from a tiny
source region to be beamed into a narrow cone, thereby amplifying the
observed luminosity without enhancing the energy requirement of the
emitting region. This is the key idea of the ``jets in a jet"
model. Note that \cite{Ghisellini+09} suggest an alternative way of
reducing the size of the emitting region, which we do not consider in
this paper.

In models of magnetically accelerated jets, during the acceleration
phase the radius and Lorentz factor of the jet generally scale as
$R\sim \Gb^2 R_g$ (e.g., \citealt{Tch+08}). This gives a causal scale
$R' = R/\Gb \sim \Gb R_g$.  If acceleration ceases and the radiation
is produced when the jet is in a coasting phase, then $R'>\Gb R_g$.
To allow for both possibilities, we write:
\begin{equation}
R = f_c \Gb^2 R_g, \qquad R' = \frac{R}{\Gb} = f_c \Gb R_g,
\label{eq:R}
\end{equation}
where the numerical factor $f_c \ge 1$ allows for coasting.  In the
fireball model of GRBs, during acceleration we have $R\sim \Gb R_g$
and it would appear that we could have $f_c<1$. However, energy
dissipation in a baryon-loaded fireball typically occurs ony at a
distance $R \approx \Gb^2 c\ t_{\rm engine}$, or equivalently, $t_{\rm
dyn} \approx t_{\rm engine}$, where $t_{\rm engine}$ is the
characteristic time scale of the central engine
\citep{Sari_Piran97}. Clearly $t_{\rm engine} \ge T_g$, and thus once
again we obtain equation (\ref{eq:R}) with $f_c \geq 1$.

Events such as the bright flare observed in PKS 2155-304 by
\citet{Aharonian+07} are clearly rare.  Most of the time the source is
much fainter.  To model this behavior we assume that the source has
more or less steady low-level jet activity, but occasionally goes
through short periods of time when the engine power becomes very much
larger.  We are concerned with the properties of these bursts of more
energetic activity; one of these bursts presumably produced the bright
flare seen in PKS 2155-304. Let $t_{\rm engine}$ be the duration that
the source spends in an energetic state.
Clearly $t_{\rm engine} \ge T_g$. In addition, if $t_{\rm engine}$ is
very small, radial spreading will cause the shell to expand to a width
$\sim R/\Gb^2$, as discussed earlier. We thus write the radial
width of the emitting region as
\begin{equation}
\Delta R = {\rm max}\left(ct_{\rm engine},\frac{R}{\Gb^2}\right)\equiv
f_d \frac{R}{\Gb^2} = f_d\frac{R'}{\Gb} = f_d f_c R_g, \qquad \Delta
R' = f_d R' = f_d f_c \Gb R_g,
\label{eq:DeltaR}
\end{equation} 
where the engine duration factor $f_d \ge 1$. The overall duration of
the flaring activity is determined by $\Delta R$, since the angular
spreading time and dynamical time are always shorter than or equal to
the radial time. Hence
\begin{equation} 
T \approx \frac{\Delta R}{c} = f_d \frac{R'}{\Gb c} = f_d f_c T_g.
\label{eq:T}
\end{equation}
In PKS 2155-304, $T$ is roughly equal to $T_g$ for our fiducial choice
of the BH mass ($M_9=2$). Thus the product $f_d f_c$ must be of order
unity. Since $f_d$ and $f_c$ are individually larger than or equal to
unity, this implies that each is of order unity. If we assume a lower
mass for the BH, e.g., $M_9=0.5$, then $f_d f_c\approx 4$ and we have
some freedom in choosing the values of $f_d$ and $f_c$.

As explained above, the key idea of the ``jets in a jet" model is that
the radiating fluid in the jet is sub-divided into a number of
independent volumes, each moving relativistically with respect to the
mean frame of the jet. In order to explain the rapid variability, it
is necessary for each sub-volume to be much smaller than the causality
scale $R'$, or the engine-related scale $\Delta R'$.  The breaking up
of the jet into kinematically distinct sub-volumes cannot be related
to the engine since the length scale involved is too small. Rather, it
must result from a local instability of some sort.  Such a situation
can arise naturally in a highly magnetized outflow, for instance
through magnetic reconnection or MHD turbulence. However, in the
following, we do not assume anything specific about the nature of the
outflow or what causes the instability.

There are two main variants of the ``jets in a jet" model
\citep{Lazar+09}.  The distinction is whether the emission duration of
individual pulses in the flare is determined by physics in the
comoving frame of the jet or in a frame moving relative to the jet.

\subsection{Subjets and reconnection} 

In the model described by Giannios et al. (2009; see also Lyutikov
2006, Lazar et al. 2009 in the context of GRBs), which we denote
hereafter as the ``subjets" model, magnetic field reconnection cells
arise sporadically within the strongly magnetized jet fluid. Each
reconnection event leads to the ejection of twin subjets of
relativistic plasma with a typical Lorentz factor $\Gj$ as measured in
the mean frame of the jet. The subjets emit the observed TeV emission.
A single pulse in the observed TeV lightcurve corresponds to a single
subjet.  Hence the observed duration of a pulse is equal to the time
taken for the completion of a reconnection event as measured in the
jet frame, divided by $\Gb$ (to transform to the observer frame).
While this model is strongly motivated by magnetic reconnection, there
in nothing in the following discussion that depends on the specific
details of this mechanism. Hence the model is valid for any process in
which a local intability produces relativistic subjets.

Each reconnection event dissipates the magnetic energy in a certain
characteristic volume in the jet frame.  Let $\lp$ be the typical
length scale of this volume, and let $\beta'=v'/c$ be the typical
speed with which the magnetic energy flows into the central
reconnection zone. The duration of the reconnection event, as measured
in the jet frame, is $\delta t' \approx \lp/\beta' c$, and so the
observed variability time is:
\begin{equation}
\delta t \approx  \frac{\lp}{\Gb {\beta'} c} . 
\label{eq:dt1}
\end{equation}
Clearly, if a subjet changes its direction by more than $1/\Gj$ before
consuming all the energy within the reconnection volume then the
observed pulse will be shorter\footnote{Correspondingly, the shape of
the pulse in the light curve will no longer be determined by the onset
and decline of the reconnection event, but will be determined by the
motion of the jet.}.  In this case, we simply redefine $l'$ such that
the subjet direction is constant to within $1/\Gj$ during the time
$\delta t' = \lp/\beta' c$.  Making use of equation (\ref{eq:T}), the
ratio of the flare duration to the variability time is:
\begin{equation}
\frac{T}{\delta t}  \approx f_d {\beta'} \frac{R'}{ \lp} .
\label{eq:Tdt1}
\end{equation}

For a given observer, the region from which radiation can be seen has
a size $\sim R'$ perpendicular to the line-of-sight and a size $\sim
\Delta R'=f_d R'$ (measured in the jet frame) along the line-of-sight.
Since each independent reconnection region has a volume $\sim \lp^3$,
the total number of subjets within the observed volume is
\begin{equation}
n_{\rm tot}  \approx 2f_d  \left( \frac{ R'}{ \lp} \right)^3 = 
\frac{2} {f_d^2  {\beta'}^3} \tdtpar^3.
\label{eq:ntot1}
\end{equation}
The factor of 2 is because each reconnection site produces two subjets
moving in opposite directions.  Purely from relativistic beaming, each
subjet would illuminate a solid angle $\sim \pi/\Gj^2$ in the jet
frame.  Since there might be additional beam broadening due to
intrinsic velocity fluctuations within the subjet, we write the solid
angle illuminated by one subjet as $\Omega'_j \equiv \pi f_j /\Gj^2$
with $f_j \ge 1$.  Assuming that subjets from different reconnection
regions are uncorrelated and are oriented randomly, a given observer
receives radiation from a fraction $\sim \Omega'_j/4\pi$ of the
subjets.  Each visible subjet produces one pulse in the observed
lightcurve.  We thus estimate the duty cycle in this model to be:
\begin{equation}
\xi \approx \frac{f_j}{ 2 \Gj^2 f_d^2 {\beta'}^3} \tdtpar^2 \approx
\frac{f_j}{2 \beta'}\left(\frac{R'}{\Gj\lp}\right)^2.
\label{eq:np1}
\end{equation}

Using our fiducial numbers for PKS 2155-304, viz., $T/\delta t \approx
70$ and $\xi \approx 0.4$, equation (\ref{eq:np1}) gives
\begin{equation}
\frac{\Gj^2 f_d^2 {\beta'}^3}{f_j} \approx 6000.
\end{equation}
We showed earlier that $f_d\approx 1$, and by definition we have
$f_j\geq 1$.  In addition, current understanding of relativistic
reconnection suggests that ${\beta'} \sim 0.1$ \citep{Lyubarsky05}.
Writing ${\beta'}=0.1\beta'_{-1}$, we thus find 
\begin{equation}
\Gj \approx 2500 f_j^{1/2} f_d^{-1} {\beta'^{-3/2}_{-1}}.
\end{equation}
This is an extremely large value, particularly when we recall that
\citep{Lyubarsky05} $\Gj \sim \sqrt{\sigma}$, where $\sigma$ is the
magnetization parameter \citep{Kennel_Coroniti_84}.  We require
$\sigma$ to be truly enormous, which leads to other problems (see
below).  Even if we set $\beta'=1$ ($\beta'_{-1}=10$), which is
unlikely, we obtain $\Gj \approx 80$. The subjets thus need to move
highly relativistically with respect to one other and with respect to
the mean jet frame.  Also, equation (\ref{eq:ntot1}) indicates that
there must be $n_{\rm tot} \approx 7\times 10^8{\beta'^{-3}_{-1}}$
independent subjets, which is again very extreme.

Extreme values of the parameters, e.g., $T\approx 4000$\,s (which
requires a lower BH mass $M_9<0.4$ to maintain $T\geq T_g$), $\delta t
\approx 600$\,s (the longest time scale consistent with the
observations), improve the situation somewhat.  With $\beta'=0.1$,
this gives $\Gj\approx170$, $n_{\rm tot}\approx 7\times10^5$, still
rather extreme, while with $\beta'=1$, we obtain reasonable values:
$\Gj\approx 5$, $n_{\rm tot}\approx 700$. Alternatively, we could
assume that the BH has a smaller mass than our fiducial value
$M_9=2$. For instance, if we take $M_9=0.5$ and assume $f_c=1$, we
obtain $f_d=4$.  If we further set $f_j=1$, $\beta'=1$, then $\Gj \sim
20$ and $n_{\rm tot} \approx 4\times 10^4$.

An additional issue is that, in order for relativistic reconnection to
be energetically efficient, we need the inflowing magnetic field lines on
the two sides of the reconnection region to be aligned to within an
angle $\sim 1/2\sqrt{\sigma}$ \citep{Lyubarsky05}. Recalling that $\Gj
\sim \sqrt {\sigma}$ and that the model requires a large $\Gj$,
this poses a serious problem.

\subsection{Relativistic turbulence} 

Another scenario, which was advocated by Narayan \& Kumar (2009;
discussed further by Lazar et al. 2009) and which we denote the
``turbulence'' model, is one in which relativistic turbulence is
generated, possible because  of some MHD instability in the jet
fluid. As a result, blobs of fluid move in random directions with a
typical Lorentz factor $\Gj$ as measured in the mean jet frame.  Each
blob is roughly spherical in its own frame, which means it has a
transverse size $\lp$ and longitudinal size $\lp/\Gj$ in the jet
frame.  At any instant, the radiation from a given blob is focused
into a solid angle $\sim \pi/\Gj^2$ in the jet frame.  However, over
time the beam orientation might wander, so we write the solid angle
illuminated by a blob as $\Omega'_j = \pi
f_j/\Gj^2$. \cite{Narayan_Kumar09} suggested that the velocity vector
of a blob might wander by about a radian during the time a blob
radiates, which corresponds to $f_j \sim \Gj$. However, more general
situations, including $f_j\approx 1$, can also be considered
\citep{Lazar+09}.

The duration of a pulse as measured in the jet frame\footnote{See
\citet{Lazar+09} for a discussion of additional relevant time scales.}
is given by the longitudinal size of a blob, i.e., $\delta t' \approx
\lp/\Gj c$.  Transforming to the observer frame, the observed pulse
duration is:
\begin{equation}
\delta t \approx  \frac{\lp}{\Gb \Gj c} . 
\label{eq:dt2}
\end{equation}
The ratio of the duration of the flaring event to the duration of a
single pulse is then:
\begin{equation}
\frac{T}{\delta t}  \approx f_d \Gj \frac{R'}{\lp} .
\label{eq:Tdt2}
\end{equation}

The total number of emitting regions within the observed volume is
(there is no additional factor of $2$ here since each blob radiates
into a single beam):
\begin{equation}
n_{\rm tot}  \approx f_d  \left( \frac{ R'}{ \lp} \right)^3 = \frac{1}{f_d^2  \Gj^3} \tdtpar^3 ,
\label{eq:ntot2}
\end{equation}
and the duty cycle is:
\begin{equation}
\xi \approx \frac{f_j }{ 4 \Gj^5 f_d^2 } \tdtpar^2  \approx
\frac{f_j}{4\Gj} \left(\frac{R'}{\Gj \lp}\right)^2.
\label{eq:np2}
\end{equation}
Compared to the subjets model, we see that $\beta'\to\Gj$, with
an additional factor of $2$ because of the different number of beams
per blob in the two models.

Generally, we find that the turbulence model has a much easier time
satisfying the observational constraints.  For example, with $T/\delta
t \approx 70$, $\xi \approx 0.4$, $f_j=1$, $f_d=1$, we obtain
$\Gj\approx 5$ and $n_{\rm tot} \approx 3000$, which are quite
reasonable. If $f_j \approx \Gj$ (as suggested by
\citealt{Narayan_Kumar09}), then $\Gj\approx 7$ and $n_{\rm tot}=800$,
which is again acceptable.

The huge difference in the predictions of the subjets and turbulence
models can be traced to differences in how the variability time scale
is related to the blob size $\lp$ in the two models.  In the subjets
model we have $\delta t' \approx \lp/\beta' c$, i.e., it is the
crossing time of a region of size $\lp$, as measured in the mean jet
frame, at speed $\beta' c$.  In the turbulence model, on the other
hand, we have $\delta t' \approx \lp/\Gj c$, i.e., it is the crossing
time of a region of size $\lp$, as measured in the frame of the moving
blob, at speed $c$.  The factor $\beta'\sim 0.1$ arises in the subjets
model because we believe reconnection is limited to a speed
substantially below $c$ \citep{Lyubarsky05}. There is no equivalent
factor in the turbulence model, mostly by assumption, since we do not
have as complete a physical picture of this model as we do for the
subjets model.  The second factor $\Gj$ is because of another key
difference between the two models.  In the subjets model, the time
scale for a reconnection event is determined by physics in the mean
frame of the jet, since we assume that the reconnection cell is at
rest in this frame. In the turbulence model, on the other hand,
everything is determined by physics in the comoving frame of the blob.
The net factor of $\Gj/\beta'$ between the two models is quite a large
number and this leads to drastic differences in their predictions.

\section{Opacity limits} 
\label{tau}

As discussed by \citet{Cavallo_Rees_78}, if a source of non-thermal
high energy radiation is spatially more compact than a certain limit,
then photon-photon collisions will be very frequent and there will be
copious electron-positron pair production. The source will then be
opaque to its own high energy radiation and will be unable to produce
the observed non-thermal spectrum.  Assuming a spatially homogeneous
jet, \citet{Begelman+08} estimated that the pair production optical
depth in PKS 2155-304 is
\begin{equation}
\tgg \approx 2\times 10^{10} L_{46} t_{300}^{-1} \Gb^{-6},
\label{eq:tgg1}
\end{equation}
where $L_{46}$ is the (isotropic equivalent) luminosity of the jet in
units of $10^{46}\,{\rm erg\,s^{-1}}$ and $t_{300}$ is the variability
time in units of 300\,s.\footnote{The exponenent on $\Gb$ in
eq. (\ref{eq:tgg1}) depends on the spectrum of the source.
\citet{Piran_99} gives $4+2\alpha$ instead of 6, where $\alpha$ is the
spectral index ($F_\nu \propto \nu^{-\alpha}$).  The exponent 6 thus
corresponds to $\alpha=1$, which is a reasonable value for blazars and
GRBs.}  Setting $L_{46}\approx t_{300}\approx1$, the requirement
$\tgg<1$ gives $\Gb>50$.

In the ``jets in a jet" model, the radiation received by the observer
during any given pulse in the light curve comes from a single subjet
or blob in the source. Since the emitting region moves with respect to
the observer with a net Lorentz factor $\approx \Gb\Gj$, one might be
tempted to replace $\Gb$ by $\Gb\Gj$ in equation (\ref{eq:tgg1}) to
thereby obtain $\Gb\Gj > 50$.  This would loosen the constraint on
$\Gb$ by a large factor (at least 5, and potentially several tens).
However, the argument is incorrect.\footnote{\citet{Giannios+09}
derived a bound $\Gb > 9$ for PKS 2155-304 using what appears to be,
apart from a number of details, essentially this incorrect argument.}

The result given in (\ref{eq:tgg1}) corresponds to a single radiating
region moving relativistically towards the observer.  It describes the
optical depth for the radiation to escape from its own local emission
region. Since there are many emission regions in the ``jets in a jet"
model, even after a beam escapes from its original blob, it is likely
to encounter other beams of radiation on its way towards the
observer. In order for the radiation to reach the observer, we require
the net optical depth due to all these encounters to be small.  To
calculate the corresponding optical depth, it is most convenient to
work in the mean jet frame (the primed frame in our notation).

Let us focus on the volume of the jet from which the observer can
receive radiation. This volume has a size $R'\times R'$ in the two
transverse directions and $\Delta R' = f_d R'$ in the radial
direction.  Each fluid element in this volume will radiate roughly for
a dynamical time $t_{\rm dyn} \approx R'/c$.  Multiplying this time by
$c$ to convert it to a length, the total four-volume that is visible
to the observer is
\begin{equation}
V^{(4)}_{\rm total} \approx  f_d R'^4.
\label{eq:V4tot}
\end{equation}

Consider first the subjets model.  Each subjet illuminates a conical
volume with a solid angle $f_j\pi/\Gj^2$, and so the average
three-volume of the cone is $(f_j\pi/3\Gj^2)\langle r^3\rangle$, where
$r$ is the length of the cone inside the reference volume.  For $r$
distributed uniformly between 0 and $R'$, the mean three-volume of a
subjet is $(f_j\pi/12\Gj^2)R'^3$.  The time for which a subjet shines
(as measured in the jet frame) is $\lp/\beta' c$. Multiplying by this
factor, and also by the number of subjets $n_{\rm tot}$ given in
equation (\ref{eq:ntot1}), the total four-volume occupied by all
subjets is
\begin{equation}
V^{(4)}_{\rm subjets} \approx
\frac{f_j\pi}{6\Gj^2f_d^2\beta'^4}R'^3\lp \tdtpar^3.
\end{equation}
Dividing by $V^{(4)}_{\rm total}$ and making use of equation
(\ref{eq:np1}), the fractional four-volume occupied by subjets is
\begin{equation}
f_{\rm subjets} \equiv \frac{V^{(4)}_{\rm subjets}}{V^{(4)}_{\rm total}}
\approx \frac{\pi}{3}\xi,
\end{equation}
i.e., it is roughly equal to the duty cycle of the observed light
curve, $\xi\approx 0.4$. This result is not surprising.  

A similar calculation can be done for the turbulence model. As before,
the solid angle of each beam is $f_j\pi/\Gj^2$, but the duration of
the beam in the jet frame is $\approx \lp/\Gj c$.  Repeating the same
steps as above (using eqs. \ref{eq:ntot2}, \ref{eq:np2}), we find
\begin{equation}
V^{(4)}_{\rm turbulence} \approx 
\frac{f_j\pi}{12\Gj^6f_d^2}R'^3\lp \tdtpar^3,
\end{equation}
\begin{equation}
f_{\rm turbulence} \equiv \frac{V^{(4)}_{\rm turbulence}}{V^{(4)}_{\rm
total}} \approx \frac{\pi}{3}\xi.
\end{equation}
The result is the same as for the subjets model.

Before proceeding, let us define a ``reference jet model'' which
consists of a homogeneous jet with a comoving volume $R'\times
R'\times f_d R'$, moving with a Lorentz factor $\Gb$ towards the
observer.  By construction, the total duration of the observed high
energy radiation from this hypothetical jet is the same as in PKS
2155-304: $T \approx f_d R'/\Gb c$ (see eq. \ref{eq:T}).  We assume
that the jet produces the same mean luminosity as that observed in PKS
2155-304. However, being homogeneous, it cannot produce the observed
rapid variability.

When applying equation (\ref{eq:tgg1}) to the reference jet
model, we must set the variability time equal to $R'/\Gb c = T/f_d
\approx (20000/f_d)\,{\rm s}$. As a result, the optical depth and
limiting Lorentz factor become (for $L_{46}\approx1$)
\begin{equation}
{\rm reference~jet~model}: \quad \tgg \approx 3\times10^8 f_d
\Gb^{-6}, \qquad \Gb > 25 f_d^{1/6}.
\end{equation}
The larger size of the emitting volume, compared to the model
considered by \citet{Begelman+08} where the size was constrained by
the observed variability time of 300\,s, leads to a factor of 2
reduction in the minimum bulk Lorentz factor $\Gb$.

Consider now the ``jets in a jet" model.  In both the subjets and the
turbulence versions of this model, the radiation of the criss-crossing
beams occupies a fraction $\approx \xi$ of the total four-volume
$V^{(4)}_{\rm total}$.  Since the ``jets in a jet" model produces the
same average luminosity as the reference jet model,
the mean number density of photons in the jet frame must be the same.
The only difference is that, in the ``jets in a jet" model, a fraction
$\xi$ of the volume is occupied by radiation, and in these regions the
local number density is higher than average by a factor $\xi^{-1}$.
Any given beam of radiation will intersect many other beams on its way
out\footnote{This would not be true if $\xi \ll 1$, but we are working
in a different limit where $\xi \sim 1$.}. A fraction $\xi$ of its
path will be through other beams, each of which will on average have a
photon number density a factor $\xi^{-1}$ larger than in the reference
jet model, and the remainder of its path will be through
radiation-free regions.  The net result is that each escaping photon
will on average interact with exactly the same number of high energy
photons as a photon in the reference jet model.  Thus, the optical
depth to pair production is the same in both models, i.e.,
\begin{equation}
{\rm ``jets~in~a~jet"~model}: \quad \tgg \approx 3\times10^8 f_d
\Gb^{-6}, \qquad \Gb > 25 f_d^{1/6}.
\end{equation}

Note that, while the optical depth is the same in both the reference
jet model and the ``jets in a jet,'' the latter model has the
distinction of being able to produce the rapidly varying light curve
observed in PKS 2155-304.  Interestingly, the model does this without
suffering any penalty in the pair production opacity.  In fact, the
``jets in a jet" model has a less restrictive limit on the bulk
Lorentz factor of the jet, $\Gb\,\gsim\,25$, compared to the original
limit, $\Gb\,\gsim\,50$, obtained by \citet{Begelman+08}.

\section{Implications and conclusions}
\label{implications}

The very rapid variability observed in the TeV flux of PKS 2155-304
\citep{Aharonian+07}, which is faster by a factor of 50 than the
gravitational time scale of the central BH, is a remarkable
puzzle. Normally, even assuming a relativistic outflow, one does not
expect to see variability faster than the gravitational time of the
driving engine.  PKS 2155-304 violates this expectation by a large
factor.  A possible way out --- apparently the only way out --- is to
invoke some version of the ``jets in a jet'' model.  This model was
discussed earlier (though this name was not used) in connection with
the variability of GRBs
\citep{Lyutikov_Blandford03,Lyutikov06,Narayan_Kumar09,Lazar+09}).
The same idea has been shown to explain the variability in PKS
2155-304 \citep{Giannios+09}.

In the ``jets in a jet" model the fluctuating pulses in the observed
radiation are not produced by the entire jet but by small sub-regions
within the jet. These sub-regions move relativistically with respect
to the mean frame of the jet. As a result each sub-region produces a
beam of radiation that is focused tightly in the direction of its
motion.  Only a few sub-regions radiate towards the observer, but
these appear anomalously bright as a result of relativistic beaming.
Thus, the model explains both the large apparent luminosity and the
rapid variability, without violating any energy or light-crossing-time
constraints.

We have considered in this paper two versions of the ``jets in a jet''
model. The first version, the ``subjets'' model, assumes that something like magnetic
reconnection takes place in many different sub-regions within the jet
fluid and that each reconnection site ejects twin relativistic subjets
in opposite directions along the local recombining magnetic field.
This model is characteristic of a class of models in which the time
scale of pulses in the observed light curve is determined by a process
that is at rest in the frame of the jet. The role of the relativistic
subjet is merely to provide a luminosity boost through beaming.  The
second version, the ``turbulence'' model, assumes that some
instability in the jet fluid leads to highly relativistic random
motions within the medium and as a result the radiation from each
sub-region is narrowly beamed in the direction of the local
motion. This model is characteristic of a class of models in which the
observed variability time scale is determined by a process that takes
place in the frame of the relativistically moving sub-region.  Thus,
both the observed variability time scale and the luminosity are
affected by the motion of the sub-region (in contrast to the subjets
model where only the luminosity is modified).  These two models are
quite similar in spirit, but they differ in details which become
important in the case of extreme objects such as PKS 2155-304.

In the case of the subjets model, using our canonical
observational constraints from PKS 2155-304, viz., flare duration $T
\approx 20000$\,s, variability time $\delta t \approx 300$\,s, pulse
duty cycle $\xi \approx 0.4$, reconnection inflow speed $\beta'
\approx 0.1$, we find that the model requires the Lorentz factor of
the subjets, as measured in the mean jet frame, to be $\Gj \approx
2500$. Given that $\Gj \sim \sqrt{\sigma}$, where $\sigma$ is the
magnetization parameter of the jet fluid, this value seems
unreasonably large.  Perhaps of even greater concern, the model
requires the total number of independent reconnection sites, or
sub-regions, to be $n_{\rm tot} \approx 10^9$.  Both requirements
become less stringent if we assume extreme values for the time scales,
e.g., $T \approx 4000$\,s (i.e., setting $T$ equal to the duration of
the observations, assuming that the flare started exactly when the
observations began) and $\delta t \approx 600$\,s (the maximum
duration of each pulse, see \citealt{Aharonian+07}). For this choice,
$T/\delta t \approx 7$, $\xi \approx 0.8$, and we find $\Gj \approx
170$, $n_{\rm tot} \approx 10^6$.

The crux of the problem in the subjects model is the relatively low
fiducial value we assume for $\beta'$ (based on the work of
\citealt{Lyubarsky05}).  The velocity $\beta' c$ is the speed with
which magnetized fluid moves in towards the reconnection point. Since
this velocity determines the variability time, a low value of $\beta'$
implies a correspondingly small size $\lp$ of the reconnection
cell. To compensate, $\Gj$ has to be very large so that the small
amount of energy available within one reconnection cell is strongly
beamed to give the luminosity observed in a single pulse of the light
curve.

It is possible that the arguments leading to $\beta' \approx 0.1$
could be circumvented, allowing a reconnection inflow speed close to
$c$. It is also possible that some (unknown) mechanism other than reconnection produces
the subjets and that this mechanism transports energy to the subjets
at the speed of light. However, even if we set $\beta'=1$, this is
still only marginally acceptable. Using fiducial values for the other
parameters we find $\Gj \approx 80$, $n_{\rm tot} \approx 10^6$. We
obtain a reasonable solution only if we choose both a high value of
$\beta' \approx 1$ and extreme values of the time scales, $T \approx 4000$\,s,
$\delta t \approx 600$\,s.  In this case, we find $\Gj \approx 5$,
$n_{\rm tot} \approx 10^3$.  Or, we could choose $\beta' \approx 1$ and
instead of changing $T$ and $\delta t$, we could select a smaller BH
mass, $M_9 \approx 0.5$, which gives $\Gj\approx 20$, $n_{\rm
tot}\approx 4\times 10^4$.

We thus conclude that the subjets model works only if magnetic
reconnection in relativistic plasmas proceeds at the speed of light,  much faster than currently
thought.  This alone may not be sufficient to explain PKS 2155-304.  We may also need to push the
observational estimates of time scales or BH mass to extreme values.

The situation in the case of the turbulence model is quite different.
In contrast to the subjets model, here, even with fiducial values of
parameters, the implied conditions in the jet are quite
reasonable. For instance, with $T \approx 20000$\,s, $\delta t \approx
300$\,s, $\xi \approx 0.4$, we find $\Gj\approx5-7$, $n_{\rm
tot}\approx10^3$.  Such conditions appear quite reasonable for a
relativistically moving high-energy source.

The weakness of the turbulence model is that it is short on details.
We do not have a physical model of what causes the turbulence and what
the limits of this process are. (This is in contrast to the subjets
model where we have a very specific process in mind -- reconnection --
which immediately gives a physical constraint $\beta'\sim0.1$.)  In
fact, the word ``turbulence'' itself is merely a code for random
motions. The actual dynamics may not be truly turbulent in the sense
it is normally understood.  Certainly, we cannot have hydrodynamic
turbulence since motions with $\Gj\approx 5$ are highly supersonic
with respect to the maximum allowed sound speed, $c_{\rm
s,max}=c/\sqrt{3}$.  In principle, MHD turbulence is compatible with
the model since wave speeds can reach up to Lorentz factors
$\Gamma_{\rm wave}\sim\sqrt{\sigma}$, where $\sigma$ is the
magnetization parameter.  Therefore, with a sufficiently strongly
magnetized medium ($\sigma \gg 1$), random motions with $\Gj \approx
5-7$ could be supported.

The heating of particles and the production of radiation is also
unexplained in the turbulence model.  In the subjets model,
reconnection automatically produces relativistic beams of particles
and it is not hard to imagine that these particles will radiate in the
ambient magnetic field.  (Note, however, that the dissipation
efficiency in reconnection may be rather low unless the reconnecting
fields coming in from the two sides are practically parallel, see
\citealt{Lyubarsky05}).  In the case of the turbulence model, the most
obvious candidate for particle heating is a shock, but this is
unlikely to work since highly magnetized shocks are very inefficient
at particle heating \citep{Kennel_Coroniti_84,Narayan+11}.  Thus, it
may be necessary to invoke heating through reconnection inside the
turbulent blobs (which would immediately introduce inefficiency
through a $\beta'$-like factor, just as in the subjets model).
Perhaps the most promising idea is that the particles are heated
directly by waves in the turbulent magnetized medium, but the details
remain to be worked out.

Summarizing, we find the situation far from clear. The subjets model
with magnetic reconnection provides a natural way to produce the
required ``jets within a jet.'' However, given our current understanding
of relativistic reconnection \citep{Lyubarsky05}, viz., that the
inflow speed towards the reconnection site can be no larger than a
tenth the speed of light, this model has great difficulty explaining
the observations in PKS 2155-304.  The turbulence model
\citep{Narayan_Kumar09,Lazar+09} apparently has no difficulty fitting
the data.  However, all we have is a broad outline of this model, and
there is no physical picture of how the ``turbulence'' in this model
is produced or how this turbulence heats particles and produces the
observed radiation.

In \S4 we analyze the pair production opacity for TeV photons
in the ``jets in a jet'' model.  We show that collisions of beams
emitted from different sub-regions are common and hence the opacity
due to these beam-beam collisions is more important than the opacity
within a single emitting sub-region.  Allowing for this effect we find
that, regardless of which version of the ``jets in a jet" model we
consider, the minimum bulk Lorentz factor $\Gb$ of the jet is $\approx
25$, which is lower than the limit $\approx 50$ obtained by
\citet{Begelman+08}.  Although the reduction in the value of $\Gb$ is
only a factor of 2, it is nevertheless a significant revision since
the new value is closer to the typical bulk Lorentz factors
$\approx10-20$ found in relativistic jets in other blazars.

This work was supported in part by NSF grant AST-1041590 and NASA
grant NNX11AE16G (RN), and by the Israel Center for Excellence for
High Energy Astrophysics and an ERC advanced research grant (TP).

\appendix
\numberwithin{equation}{section}
\section[A]{Bayesian estimate of the duration of the flare in 
PKS 2155-304}

Let us suppose that an astronomical source is on when we first start
observing it. It remains on for a time $\tobs$ and then shuts off. We
would like to estimate the total on-time $\ton$ of the source,
including the time it was on before we began observing.

Let us suppose that the source was observed previously a time $\tmax$
before the current turn-off time and that, at that time, the source
was off.  Therefore, the minimum and maximum possible on-time of the
source are $\tobs$ and $\tmax$, respectively. We wish to evaluate the
probability distribution of $\ton$ between these two limits.

In the absence of any other information, the probability distribution
of $\ton$ is logarithmically flat, i.e.,
\begin{equation}
P(\ton)d\ton \propto d\ln\ton = d\ton/\ton.
\end{equation}
For a given value of $\ton$, the chance that the observed on-time will
lie between $\tobs$ and $\tobs+d\tobs$ is clearly uniform for all
values of $\tobs$ between 0 and $\ton$. Thus, for a true on-time
$\ton$, the probability of an observed on-time $\tobs$ is simply
\begin{equation}
P(\tobs|\ton)d\tobs = d\tobs/\ton, \quad 0 \leq \tobs \leq \ton.
\end{equation}
This distribution is normalized such that the total integrated
probability is unity.

Bayes' theorem states that $P(\ton|\tobs)$, the probability that the
true duration is $\ton$, given an observed duration $\tobs$, is
\begin{equation}
P(\ton|\tobs) = C P(\tobs|\ton) P(\ton),
\end{equation}
where $C$ is a normalization constant. Thus
\begin{equation}
P(\ton|\tobs)d\ton = (\tobs/\ton^2)d\ton, \quad \tobs \leq \ton \leq \tmax,
\end{equation}
where the term $\tobs$ in the numerator on the right-hand side takes
care of the normalization (assuming for simplicity that $\tmax \gg
\tobs$). The probability distribution $P(\ton|\tobs)$ peaks at
$\ton=\tobs$ (which is reasonable), but it has a long tail extending
all the way to $\ton=\tmax$.  The median and mean values of $T$
are easily calculated from this probability distribution:
\begin{equation}
T_{\rm median} \approx 2 \tobs,
\end{equation}
\begin{equation}
T_{\rm mean} = \ln(\tmax/\tobs)\,\tobs.
\end{equation}

For the particular flare of interest in PKS 2155-304, we have $\tobs
\approx 4000$\,s, so $T_{\rm median} \approx 8000$\,s.  To compute the
mean we need to know $\tmax$.  If the source was observed the night
previous to the flare and it was off at that time, then $T_{\rm max}
\sim 10^5$\,s, and we obtain $T_{\rm mean} \approx 3\tobs \approx
12000$\,s.  However, if the previous observation in the off state was
as long ago as Sept. 2003 \citep{Aharonian+05}, then $T_{\rm mean}
\sim 10\tobs \approx 40000$\,s.

\section[B]{ Comparison of notations in different works}

Since the present paper draws on a number of earlier works, notably
\citet{Narayan_Kumar09}, \citet{Lazar+09} and \citet{Giannios+09}, we
compare the notations in Table 1.

\begin{table*}
\begin{minipage}{126mm}
\caption{Comparison of notations}
\begin{center}
\begin{tabular}{|c|c|c|c|c|}
\hline
 & This work & Narayan \& & Lazar & Giannios \\ 
 & & Kumar (2009) & et al (2009) & et al (2009)  \\ \hline
Radial distance from the BH &    $R$ &  $R$ & $R,~R_0$ &$R=r R_g$ \\ \hline
Overall flare duration &$T$ &  $t_{\rm burst} $ & $T$ &  - \\ \hline
Individual pulse time scale &    $\delta t$ &  $t_{\rm var} $ & $\delta t$ &$t_f$ \\ \hline
Duty cycle & $\xi $  & $\approx 1$ & $n_p$ &  - \\ \hline
Bulk Lorentz factor &  $ \Gb$ & $\Gamma$ & $\Gamma$ &$\Gamma_j$  \\ \hline
(Flare duration)/($R/2 \Gb^2 c$) & $f_d$ &  $1$ &  $d $ & -  \\ \hline
Random Lorentz factor & $\Gj$ & $\gamma_t$& $\gamma'$ & $\Gamma_{\rm co}$ \\ \hline
Reconnection region size & $\lp$  & - & $\lp \equiv \psi_{\rm SJ}  R $  & $\tilde l$ \\ \hline
Turbulent blob size & $\lp$ & $r_e$ &  $\lp \equiv \psi  R $ & - \\ \hline
Solid angle of each beam  &$\Omega'_j \equiv \pi f_j /\Gj^2$ & $1/\gamma_t$  &  $\phi^2$  & $1/4 \Gamma_{\rm co}^2$ \\ \hline

\end{tabular}
\end{center}
\end{minipage}
\label{default}
\end{table*}

\bibliography{ms}

\end{document}